\documentclass[%
 reprint,
 amsmath,amssymb,
prl,
]{revtex4-1}

\usepackage{placeins}

\usepackage{amssymb} 

\usepackage{paralist}
\usepackage{atlasphysics}
\usepackage{subfigure}
\usepackage{multirow} 
\usepackage{graphicx}
\usepackage{dcolumn}
\usepackage{bm}
\usepackage[hyperindex,breaklinks]{hyperref}

\usepackage{amsmath} 

\usepackage{natbib}

\newcommand\progname[1]{{\sc #1}}

\newcommand{\papertitle}{Search for Higgs and $Z$ Boson Decays to $\Jpsi\,\gamma$ and $\Upsilon(nS)\,\gamma$ with the ATLAS Detector}

\newcommand{\papertitleWithBreaks}{Search for Higgs and $Z$ Boson Decays\\ to $\Jpsi\,\gamma$ and $\Upsilon(nS)\,\gamma$ with the ATLAS Detector}

\newcommand{\paperabstract}{A search for the decays of the Higgs and
$Z$ bosons to $\Jpsi\,\gamma$ and $\Upsilon(nS)\,\gamma$ ($n=1,2,3$)
is performed with $pp$ collision data samples corresponding to
integrated luminosities of up to $20.3\,\ifb$ collected at
$\rts=8\TeV$ with the ATLAS detector at the CERN Large Hadron
Collider. No significant excess of events is observed above expected
backgrounds and 95\% CL upper limits are placed on the branching
fractions. In the $\Jpsi\,\gamma$ final state the limits are
$1.5\times10^{-3}$ and $2.6\times10^{-6}$ for the Higgs and $Z$
boson decays, respectively, while in the $\Upsilon(1S,2S,3S)\,\gamma$
final states the limits are $(1.3,1.9,1.3)\times10^{-3}$ and
$(3.4,6.5,5.4)\times10^{-6}$, respectively.}

\usepackage{preprintcover}
\PreprintCoverPaperTitle{\papertitle}
\PreprintIdNumber{CERN-PH-EP-2014-294}  
\PreprintCoverAbstract{\paperabstract}
\PreprintJournalName{Physical Review Letters}

\begin{document}

\clearpage

\title{\papertitleWithBreaks}
\author{G. Aad \it{et al.}}
\collaboration{ATLAS Collaboration}
\address{}

\date{\today}

\begin{abstract}
\paperabstract
\end{abstract}

\maketitle

Rare decays of the recently discovered Higgs
boson~\cite{ATLAS:2012af,Chatrchyan:2012ia} to a quarkonium state and a photon may offer unique sensitivity to both the magnitude and sign of
the Yukawa couplings of the Higgs boson to
quarks~\cite{VatoHBranching,HBranching,Bodwin:2014bpa,Kagan:2014ila}. These
couplings are challenging to access in hadron colliders through the direct $H\to
q\bar{q}$ decays, owing to the overwhelming QCD background~\cite{Delaunay:2013pja}.

Among the channels proposed as probes of the light quark Yukawa couplings~\cite{HBranching,Kagan:2014ila}, those with the
heavy quarkonia $\Jpsi$ or $\Upsilon(nS)$ ($n=1,2,3$), collectively denoted as ${\cal Q}$, in the final state are the most readily
accessible, without requirements for dedicated triggers and
reconstruction methods beyond those used for identifying the $\Jpsi$ or $\Upsilon$. 
In particular, the decay $H\to\Jpsi\,\gamma$ may represent a
viable probe of the $H\ccbar$ coupling~\cite{HBranching}, which is sensitive to physics beyond the Standard Model (SM)~\cite{Giudice:2008uua,Kagan:2009bn}, at the Large Hadron Collider (LHC). The expected SM branching fractions for these decays have been calculated to be
${\cal B}\left(H\to\Jpsi\,\gamma\right)=(2.8\pm0.2)\times 10^{-6}$,
${\cal B}\left[H\to\Upsilon(nS)\,\gamma\right]=\left(6.1^{+17.4}_{-6.1},2.0^{+1.9}_{-1.3},2.4^{+1.8}_{-1.3}\right)\times10^{-10}$~\cite{Bodwin:2014bpa}. No experimental information on these 
branching fractions exists. These decays are a source of background and potential control sample for the nonresonant decays $H\to\mumu\gamma$. These nonresonant decays are sensitive to new physics~\cite{Bhattacharya:2014rra}.

Rare decay modes of the $Z$ boson have attracted attention
focused on establishing their sensitivity to new
physics~\cite{Perez:2003ad}. Several estimates of the SM branching fraction for the decay
$Z\to\Jpsi\,\gamma$ are available~\cite{Guberina:1980dc,ZBranching,petriello} with the most
recent being $\left(9.96\pm 1.86\right)\times10^{-8}$~\cite{petriello}. Measuring these
$Z\to \cal{\cal Q}\,\gamma$ branching fractions, benefiting from the larger production cross section relative to the Higgs case, would provide an important 
benchmark for the search and eventual observation of $H\to{\cal Q}\,\gamma$ 
decays. Additionally, experimental access to resonant ${\cal Q}\,\gamma$ decay modes 
would also provide an invaluable tool for the more challenging measurement of 
inclusive associated ${\cal Q}\,\gamma$ production, which has been suggested as a 
promising probe of the nature of quarkonium production in hadronic collisions~\cite{Lansberg:2009db,Li:2014ava}.

The decays $Z\to{\cal Q}\,\gamma$ have not yet been observed, with the only experimental
information arising from inclusive measurements, such as ${\cal  B}\left(Z\to\Jpsi X\right) =
\left(3.51^{+0.23}_{-0.25}\right)\times10^{-3}$ and the 95\%
confidence level (CL) upper limits ${\cal B}\left[Z\to\Upsilon(nS) X\right] < (4.4,13.9,9.4)\times10^{-5}$, from LEP
experiments~\cite{PDG,Acciarri:1998iy, Alexander:1995qb,Abreu:1994rk,Alexander:1995vh}.

This Letter presents a search for decays of the recently observed Higgs
boson and the $Z$ boson to $\Jpsi\,\gamma$ and $\Upsilon(nS)\,\gamma$
final states. The decays $\Jmumu$ and $\Upsilon(nS)\to\mumu$ are used
to reconstruct the quarkonium states. The search is performed with a
sample of $pp$ collision data corresponding to an integrated
luminosity of $19.2\,\ifb$ ($20.3\,\ifb$) for the $\Jpsi\,\gamma$
$\left[\Upsilon(nS)\,\gamma\right]$ analysis respectively, recorded at a center-of-mass energy
$\rts=8\TeV$ with the ATLAS detector~\footnote{ATLAS uses a right-handed coordinate 
system with its origin at the nominal interaction point (IP) in the center of the detector
 and the $z$-axis along the beam pipe. The $x$-axis points from the IP to the center of the 
 LHC ring, and the $y$-axis points upward. Cylindrical coordinates $(r,\phi)$ are used
  in the transverse plane, $\phi$ being the azimuthal angle around the
  beam pipe. The pseudorapidity is defined in terms of the polar angle
  $\theta$ as $\eta=-\ln\tan(\theta/2)$.}, described in detail in
Ref.~\cite{Aad:2008zzm}.

Higgs~boson production is modeled using the \progname{POWHEG-BOX} Monte
Carlo (MC) event
generator~\cite{powheg1,powheg2,powheg3,powheg4,powheg5}, separately
for the gluon fusion (ggF) and vector-boson fusion (VBF)
processes calculated in Quantum Chromodynamics (QCD) up to next-to-leading order in $\alphas$.
The Higgs~boson transverse momentum (\pt)
distribution predicted for the ggF process is reweighted to match the calculations of
Refs~\cite{deFlorian:2012mx, Grazzini:2013mca}, which include QCD
corrections up to next-to-next-to-leading order and QCD soft-gluon resummations up 
to next-to-next-to-leading logarithms. Quark mass effects in ggF production~\cite{Bagnaschi:2011tu} are also accounted
for.

Physics beyond the SM that modifies the charm coupling can also change production dynamics and branching fractions. In this analysis we assume the production rates and dynamics for a SM Higgs boson with $\mH = 125\:\gev$, obtained 
from Ref.~\cite{Heinemeyer:2013tqa}, with an uncertainty on the dominant ggF production mode of $12\%$. 
The VBF signal model is appropriately scaled to account for the production of a Higgs~boson in association with a $W$ or
$Z$~boson or in association with a $t\bar{t}$ pair, correcting for the relative production rates and
experimental acceptances for these channels.
Contributions from nonresonant
$H\to(Z^*/\gamma^*)\gamma\to\mumu\gamma$ decays are expected to be
negligible with respect to the present sensitivity~\cite{Firan:2007tp,Sun:2013cba,Korchin:2014kha}.

The \progname{POWHEG-BOX} MC event generator is also used to model $Z$ boson production. 
The total cross section is estimated from
Ref.~\cite{Anastasiou:2003ds}, with an uncertainty of $4\%$.

The Higgs and $Z$~boson decays are simulated as a cascade of two-body
decays. Effects of the helicity of the quarkonium states on the dimuon
kinematics are accounted for in both cases.
For Higgs and $Z$~boson events generated using \progname{POWHEG-BOX}, 
\progname{Pythia\,8.1}~\cite{pythia,pythia81} is used to simulate showering and hadronization while
\progname{Photos}~\cite{Golonka:2005pn,Davidson:2010ew} is used to provide QED 
radiative corrections to the final state. The
simulated events are passed through the full
GEANT4 simulation of the ATLAS detector~\cite{GEANT4,2010wqa} and processed
with the same software used to reconstruct data events.

The data used to perform the search in the $\Jpsi\,\gamma$ channel
were collected using a trigger that required at least one muon
with $\pt>18\:\gev$. The events used in the $\Upsilon(nS)\,\gamma$
channel were collected with a trigger requiring an isolation muon with
$\pt>24\:\gev$ and a dimuon trigger with $\pt$ thresholds of $18$ and
$8\:\gev$ for each of the muons, respectively. Events are retained for
analysis if they were collected under stable LHC beam conditions and
the detector components were operating normally.

Muons are reconstructed from inner-detector tracks combined with
independent muon spectrometer tracks or track
segments~\cite{Aad:2014rra} and are required to have $\ptmu>3\GeV$ and
pseudorapidity $|\etamu|<2.5$. Candidate ${\cal Q}\to\mumu$ decays are
reconstructed from pairs of oppositely charged muons consistent with
originating from a common vertex.  The highest-$\pt$ muon in a pair,
called the leading muon in the following, is required to have
$\ptmu>20\gev$. Dimuons with a mass, $m_{\mm}$, within $0.2\GeV$ of
the $\Jpsi$ mass~\cite{PDG} are identified as $\Jmumu$ candidates.  In
case both muons in the pair are within $|\etamu|<1.05$, the said
requirement is tightened to $0.15\GeV$.  Dimuons with
$8.0<m_{\mm}<12.0\GeV$ are considered as $\Upsilon(nS)\to\mumu$
candidates. The transverse momentum of each ${\cal Q}\to\mumu$
candidate, $\ptmumu$, is required to exceed $36\GeV$.

Selected ${\cal Q}\to\mumu$ candidates are subjected to isolation and
vertex quality requirements. The sum of the $\pt$ of the reconstructed
inner-detector tracks and calorimeter energy deposits within $\Delta
R=\sqrt{(\Delta\phi)^{2} + (\Delta\eta)^{2}} = 0.2$ of the leading
muon is required to be less than 10\% of the muon's $\pt$. The
transverse momentum of the inner-detector track associated with the
leading muon is subtracted from the sum and the subleading muon is also subtracted if it falls within the isolation
cone. To reject backgrounds from $b$-hadron decays, the measured
transverse decay length $L_{xy}$ between the dimuon vertex and the
primary $pp$ vertex is required to be less than three times its uncertainty $\sigma_{L_{xy}}$. In this case, the primary $pp$ vertex is defined as the reconstructed vertex with the highest
$\sum_{i}p^{2}_{\mathrm{T}i}$ of all associated tracks used to form the vertex. 

Photon reconstruction is seeded by clusters of energy in the
electromagnetic calorimeter. Clusters without matching tracks are
classified as unconverted photon candidates. Clusters matched to
tracks consistent with the hypothesis of a photon conversion into an $\ee$
pair are classified as converted photon
candidates~\cite{ATL-PHYS-PUB-2011-007}. 
Reconstructed photon candidates are required to have transverse momentum $\ptgamma>36\GeV$,
pseudorapidity $|\etagamma|<2.37$, excluding the barrel/endcap
calorimeter transition region $1.37<|\etagamma|<1.52$, and to satisfy
the ``tight'' photon identification criteria~\cite{Aad:2010sp}. To further suppress the contamination from
jets, an isolation requirement is imposed. The sum of the transverse momentum of all tracks and calorimeter energy
deposits within $\Delta R = 0.2$ of the photon direction, excluding those associated with the reconstructed photon, is required to be less than 8\% of the photon's transverse momentum.

Combinations of a ${\cal Q}\to\mumu$ candidate and a photon,
satisfying $\Delta\phi(\mumu,\gamma) > 0.5$, are retained for further analysis.
To improve the sensitivity of the search,  the events are
classified into four exclusive categories, based upon
the pseudorapidity of the muons and the photon reconstruction
classification. 
Events where both muons are within the region
$|\etamu|<1.05$ and the photon is (is not) classified as a conversion
constitute the ``barrel converted'' \texttt{BC} (``barrel unconverted'' \texttt{BU})
category. Events where at least one of the muons is outside the region
$|\etamu|<1.05$ and the photon is (is not) classified as a conversion
constitute the ``endcap converted'' \texttt{EC} (``endcap unconverted'' \texttt{EU}) category. 
The number of candidates observed in each category following the complete event
 selection is shown in Table~\ref{tab:events}.
 
The total signal efficiency (kinematic acceptance, trigger and reconstruction efficiencies)  
in the $\Jpsi\,\gamma$ final state is
22\% and 12\% for the Higgs and $Z$~boson decays, respectively. The
corresponding efficiencies for the $\Upsilon(nS)\,\gamma$ final state
are 28\% and 15\%. The $m_{\mm\gamma}$ resolution is
similar for both the Higgs and $Z$~boson decays and varies between 1.2\% and
1.8\%. The $m_{\mm}$ resolution is 1.4\% and 2.4\% for the
barrel and endcap categories, respectively.

The main source of background, referred to as inclusive QCD  background, is dominated by inclusive quarkonium production where a
jet in the event is reconstructed as a photon.  For the
$\Upsilon(nS)\,\gamma$ final state, events containing $Z\to\mumu$
decays with final-state photon radiation (FSR) constitute a second
source of background, a contribution which is found to be negligible in the
$\Jpsi\,\gamma$ final state.  The normalization of both of these
background sources is extracted directly from a fit to data. The
modeling of the inclusive QCD background shape, obtained with a
data-driven approach, and of the $Z\to\mumu$ background shape,
obtained from simulation, is described in the following two paragraphs. 

The background from inclusive QCD processes is modeled with a
nonparametric data-driven approach using templates to describe the
kinematic distributions. The approach
exploits a sample of loosely selected $\mumu\gamma$ events, around
2400 in the $\Jpsi\,\gamma$ channel and around 3200 in the
$\Upsilon(nS)\,\gamma$ channel. These control samples are formed from
events satisfying the nominal ${\cal Q}\,\gamma$ selection, but with
relaxed dimuon and photon transverse momenta ($\ptgamma> 25\GeV$ and
$\ptmumu>25\GeV$) and isolation requirements (separate fractional
calorimeter energy and track momentum isolation for the photon and
dimuon system of less than 60\%). Contamination of this
sample from signal events is expected to be negligible. Probability
density functions (pdfs) used to model the $\ptmumu$, $\ptgamma$,
$\Delta\eta(\mumu,\gamma)$ and $\Delta\phi(\mumu,\gamma)$
distributions of this control sample, independently for each category,
are constructed using Gaussian kernel density
estimation~\cite{Cranmer:2000du}. To account for kinematic
correlations, the distributions of $\ptgamma$,
$\Delta\eta(\mumu,\gamma)$ and $\Delta\phi(\mumu,\gamma)$ are
estimated in eight exclusive regions of $\ptmumu$. In the case of the
dimuon and photon isolation variables, correlations are accounted for
by using two-dimensional histograms derived in five exclusive regions
of $\ptmumu$. The $m_\mm$ distributions are modeled using Gaussian
pdfs, with parameters derived from a fit to the control sample. In the
$\Upsilon(nS)\,\gamma$ channel, the data control sample is corrected
for contamination from $Z\to\mumu\gamma$ decays.  The pdfs of these
kinematic and isolation variables are sampled to generate an ensemble
of pseudocandidates, each with a complete ${\cal Q}\,\gamma$
four-vector and an associated pair of correlated dimuon and photon
isolation values. The nominal selection requirements are imposed on
the ensemble and the surviving pseudocandidates are used to
construct templates for the kinematic distributions, notably the inclusive QCD
background $m_{\mm\gamma}$ and $\ptmumu$ distributions.

The background from $Z\to\mumu\gamma$ decays is modeled with
templates derived from a sample of simulated $Z$~boson events with
$m_\mm$ in the $\Upsilon(nS)$ mass region.
To validate this background model with data, the 
sidebands of the $m_{\mm\gamma}$ distribution in several validation regions, defined
by relaxed kinematic or isolation requirements, are used to compare 
the prediction of the background model with the data. 
Good agreement within the statistical uncertainties is observed.

The composition of the inclusive QCD background and the
$Z\to\mumu\gamma$ decay contribution is investigated with data. 
The details of the composition do not enter directly the background estimation for this search, 
but the composition itself is a crucial input in feasibility studies for future searches or measurements, 
where projections of these backgrounds 
to different center-of-mass energies or luminosity conditions are needed. To facilitate this
study, the selection requirements on $m_\mm$ and
$|L_{xy}/\sigma_{L_{xy}}|$ are relaxed to include the sideband
regions. In the $J/\psi\,\gamma$ final state, a simultaneous unbinned
maximum likelihood fit to the $m_\mm$ and $|L_{xy}/\sigma_{L_{xy}}|$
distributions is performed. Once the simultaneous fit is performed,
the composition of the subset of events satisfying the nominal $m_\mm$
and $|L_{xy}/\sigma_{L_{xy}}|$ requirements is estimated. After the
complete event selection, around 56\% of the events originate from
prompt $J/\psi$ production, 3\% from non-prompt $J/\psi$ production
(from $b$-hadron decays) and 41\% are combinatoric backgrounds from
nonresonant dimuon events.

A separate simultaneous fit to the $m_{\mm\gamma}$ and $m_{\mm}$
distributions of the same sample of candidate $\Jpsi$ events finds no
significant contribution from $Z\to\mumu\gamma$ decays, a conclusion
that is also supported by a study based on simulated $Z\to\mumu$
events.

For the $\Upsilon(nS)\,\gamma$ final state a simultaneous fit is
performed to the $m_{\mm\gamma}$ and $m_\mm$ distributions. After the
full event selection, inclusive $\Upsilon(nS)$ production accounts for
7\% of events, 27\% of the events are produced in $Z\to\mumu\gamma$
decays, and 66\% of the events are associated with combinatoric
backgrounds from nonresonant dimuon events. The contribution from
$Z\to\mumu\gamma$ decays is in agreement with the MC expectation.

\begin{table}[h]
\centering
\caption{The number of observed events in each analysis category. For
  comparison, the expected background yield is given in parentheses
  for the two $m_{\mm\gamma}$ ranges of interest. The Higgs and $Z$~boson 
  contributions expected for branching fraction values of $10^{-3}$ and $10^{-6}$,
   respectively, are also shown. For $\Upsilon(nS)\,\gamma$, the $1S,2S$ and $3S$ contributions are
  summed. \label{tab:events}}
\begin{tabular}{lcrlrlcc}
\hline\hline
\parbox[t]{2mm}{\multirow{3}{*}{\rotatebox[origin=c]{90}{Category}}}  & \multicolumn{5}{c}{Observed (Expected Background)} &  \multicolumn{2}{c}{Signal} \\ \cline{2-8}
  & &\multicolumn{4}{c}{Mass Range [GeV]} &$Z$ & H \\ 
  & All & \multicolumn{2}{c}{$80$--$100$} & \multicolumn{2}{c}{$115$--$135$} & $\mathcal{B}$ $[10^{-6}]$ &  $\mathcal{B}$ $[10^{-3}]$\\\hline
\multicolumn{8}{c}{$\Jpsi\,\gamma$}\\\hline
\texttt{BU} & 30 & 9 &(8.9$\pm$1.3) & 5 &(5.0$\pm$0.9)  &  1.29$\pm$0.07  & 1.96$\pm$0.24 \\
\texttt{BC} & 29 & 8 &(6.0$\pm$0.7) & 3 &(5.5$\pm$0.6)  &  0.63$\pm$0.03  & 1.06$\pm$0.13 \\
\texttt{EU} & 35 & 8 &(8.7$\pm$1.0) & 10 &(5.8$\pm$0.8) &  1.37$\pm$0.07  & 1.47$\pm$0.18\\
\texttt{EC} & 23 & 6 &(5.6$\pm$0.7) & 2 &(3.0$\pm$0.4)  &  0.99$\pm$0.05  & 0.93$\pm$0.12 \\
\hline
\multicolumn{8}{c}{$\Upsilon(nS)\,\gamma$}\\
\hline
\texttt{BU} & 93  & 42 &(39$\pm$6) & 16 &(12.9$\pm$2.0) & 1.67$\pm$0.09 & 2.6$\pm$0.3 \\
\texttt{BC} & 71  & 32 &(27.7$\pm$2.4) & 5  &(9.7$\pm$1.2)  & 0.79$\pm$0.04 & 1.45$\pm$0.18 \\
\texttt{EU} & 125 & 49 &(47$\pm$6) & 16 &(17.8$\pm$2.4) & 2.24$\pm$0.12 & 2.5$\pm$0.3 \\
\texttt{EC} & 85  & 31 &(31$\pm$5) & 18 &(12.3$\pm$1.9) & 1.55$\pm$0.08 & 1.60$\pm$0.20 \\
\hline\hline
\end{tabular}
\end{table}

Trigger efficiencies and efficiencies for muon and photon identification 
are determined from samples of $Z\to\ell\ell$,
$Z\to\ell\ell \gamma$ ($\ell = e, \mu$) and $\Jmumu$ decays in
data~\cite{Aad:2014rra,ATLAS-CONF-2012-123}.  The systematic
uncertainty on the expected signal yield associated with the trigger
efficiency is estimated to be 1.7\%.  The photon (both converted and
unconverted) and muon reconstruction and identification efficiency
uncertainties are estimated to be 0.5\% (0.7\%) and 0.4\% (0.4\%) for
the Higgs~boson ($Z$~boson) signal, respectively. An uncertainty on
the integrated luminosity of $2.8\%$ is derived using the method
 described in Ref.~\cite{Aad:2013ucp}. The photon energy
scale uncertainty, determined from $Z\to\ee$ and validated using
$Z\to\ell\ell \gamma$ decays~\cite{ref:run1-egamma-calib}, is
propagated through the simulated signal samples as a function of
$\etagamma$ and $\ptgamma$.  The uncertainty associated with the
description of the photon energy scale in the simulation is found to
be less than 0.2\% of the three-body invariant mass while the
uncertainty associated with the photon energy resolution is found to
be negligible relative to the overall three-body invariant mass
resolution.  Similarly, 
the systematic uncertainty associated with the muon momentum measurement is 
determined using data samples of $\Jmumu$ and $Z\to\mumu$ decays 
and validated using $\Upsilon(nS)\to\mumu$ decays~\cite{Aad:2014rra}.
For the \pt~range relevant to this
analysis, the systematic uncertainties associated with the muon
momentum scale are negligible.

The uncertainty in the shape of the inclusive QCD background is
estimated through the study of variations in the background modeling
procedure. The shape of the pdf is allowed to vary around the nominal
shape within an envelope associated with shifts in the $\ptmumu$ and
$\ptgamma$ distributions. Furthermore, a separate background model,
generated without removing the contamination from $Z\to\mumu\gamma$
decays, provides an upper bound on potential mismodeling associated
with this process.

Results are extracted by means of a simultaneous
unbinned maximum likelihood fit, performed to the selected events with
$30\:\gev <m_{\mm\gamma}<230\:\gev$ separately in each of the analysis
categories. In the $\Jpsi\,\gamma$ final state, the fit is performed
on the $m_{\mm\gamma}$ and $\ptmumugamma$ distributions, while for the
$\Upsilon(nS)\,\gamma$ candidates a similar fit is performed using the
$m_{\mm\gamma}$, $\ptmumugamma$, and $m_{\mm}$ distributions. The
latter distribution provides discrimination between the three
$\Upsilon(nS)$ states and constrains the $Z\to\mumu\gamma$ background
normalization. No significant $Z\to{\cal Q}\,\gamma$ or $H\to{\cal
  Q}\,\gamma$ signals are observed, as shown in
Figs.~\ref{fig:JpsiFit} and~\ref{fig:UpsiFit}.\\
\onecolumngrid
\begin{center}
\begin{figure}[h]
\centering
\includegraphics[width=0.66\linewidth]{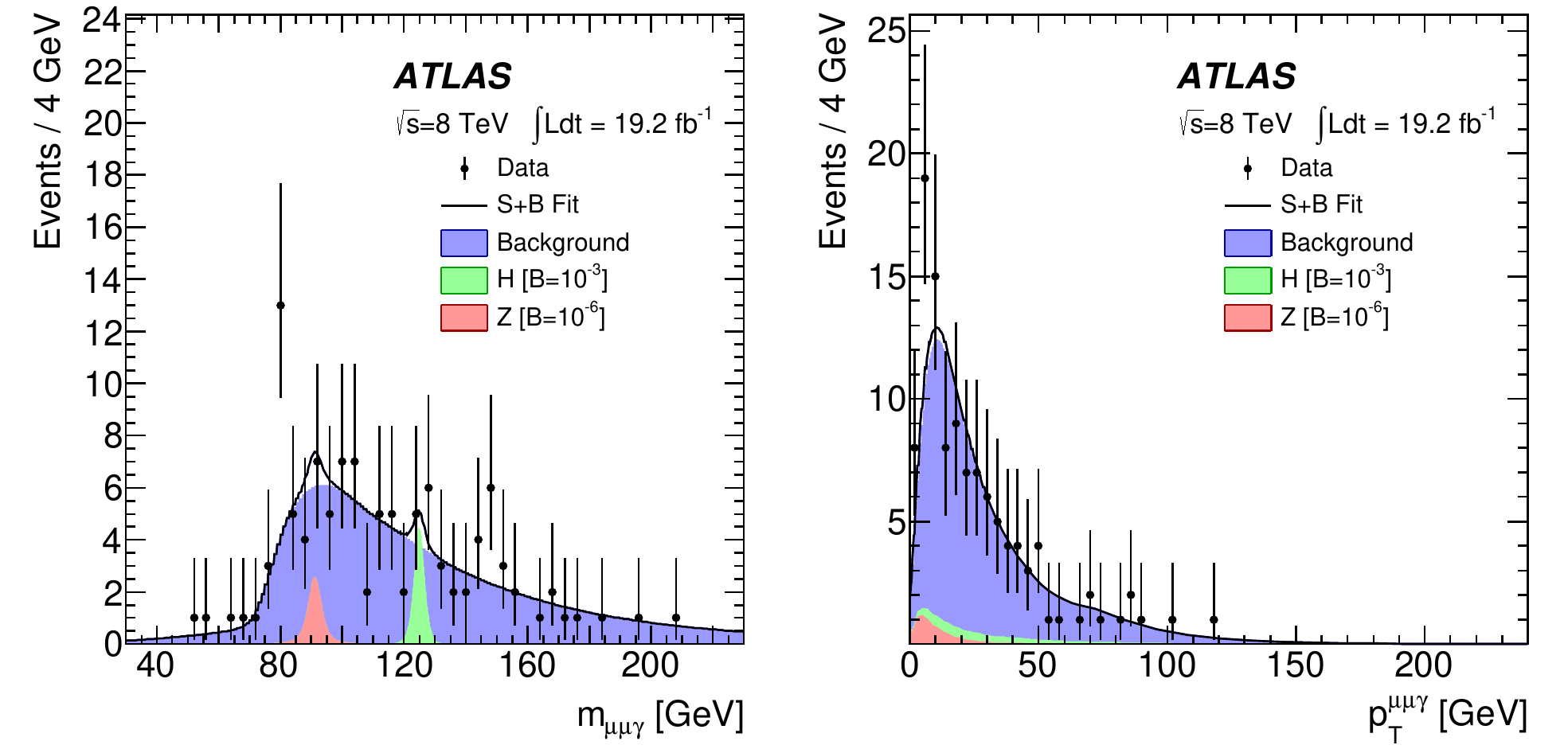}
\caption{(color online) The $m_{\mm\gamma}$ and $p_{\rm T}^{\mm\gamma}$ distributions of the selected $\Jpsi\,\gamma$ candidates, along with the results of the unbinned maximum likelihood fit to the signal and background model (S+B fit). The error bars on the data points correspond to the statistical uncertainties. The Higgs and $Z$~boson contributions as expected for branching fraction values of $10^{-3}$ and $10^{-6}$, respectively, are also shown.
\label{fig:JpsiFit}}
\end{figure}
\begin{figure}[h]
\includegraphics[width=1.0\linewidth]{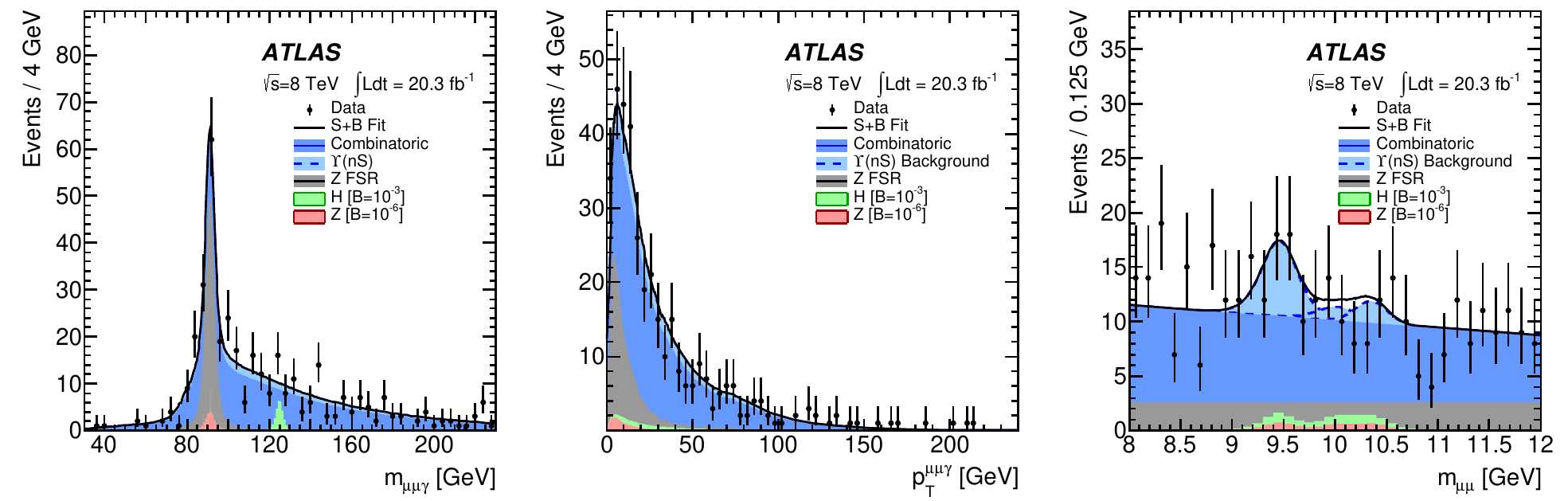}
\caption{(color online) The $m_{\mm\gamma}$, $p_{\rm T}^{\mm\gamma}$ and $m_{\mm}$ distributions of the selected $\Upsilon(nS)\,\gamma$ candidates, along with the results of the unbinned maximum likelihood fit to the signal and background model (S+B fit). The error bars on the data points correspond to the statistical uncertainties. The Higgs and $Z$~boson contributions as expected for branching fraction values of $10^{-3}$ and $10^{-6}$, respectively, for each of the $\Upsilon(nS)$ are also shown.
\label{fig:UpsiFit}}
\end{figure}
\end{center}

\twocolumngrid

Upper limits on the branching fractions for the Higgs and
$Z$~boson decays to $\Jpsi\,\gamma$ and $\Upsilon(nS)\,\gamma$
are set using the $CL_{s}$ modified frequentist formalism~\cite{cls} with the
profile likelihood ratio test statistic~\cite{Cowan:2010js}. The
expected SM production cross sections are assumed for the Higgs and
$Z$ bosons. The results are summarized in
Table~\ref{fig:Limit_Jpsi_exp}.  

The 95\% CL upper limit on the branching fraction for
$H\to\Jpsi\,\gamma$ decays corresponds to about $540$ times the
expected SM branching fraction. The upper limits on the
$Z\to\Jpsi\,\gamma$ and $Z\to\Upsilon(nS)\,\gamma$ branching fractions
significantly constrain the allowed range of values obtained from
theoretical calculations~\cite{Guberina:1980dc,ZBranching,petriello}. Upper limits are also set on the combined
branching fractions ${\cal B}\left[H\to\Upsilon(nS)\gamma\right] <
2.0\times10^{-3}$ and ${\cal B}\left[Z\to\Upsilon(nS)\gamma\right]
<7.9\times10^{-6}$, where the relative contribution of each final
state to the potential signal is profiled (allowed to float to the
values that maximize the likelihood) during the fit.

\begin{table}[h!]
\centering
\caption{Expected and observed branching fraction limits at 95\% CL for $\sqrt{s}=8\:\tev$. 
The $\pm 1\sigma$ fluctuations of the expected limits are also given. For the Higgs decay search, limits are also set on the cross section times branching fraction $\sigma\left(pp\to H\right)\times{\cal B}\left(H\to {\cal Q}\,\gamma\right)$.
\label{fig:Limit_Jpsi_exp}}
\begin{tabular}{c|c|c|c|c|c}
\hline\hline
& \multicolumn{5}{c}{95\% CL Upper Limits}\\ \cline{2-6}
 & $\Jpsi$ & $\Upsilon(1S)$ & $\Upsilon(2S)$  & $\Upsilon(3S)$ & $\sum_n \Upsilon(nS)$ \\ \hline
\multicolumn{6}{c}{${\cal B}\left(Z\to {\cal Q}\,\gamma\right)[\tabscript{10}{-6}{}]$} \\ \hline
Expected & $\tabscript{2.0}{+1.0}{-0.6}$ & $\tabscript{4.9}{+2.5}{-1.4}$ & $\tabscript{6.2}{+3.2}{-1.8}$ & $\tabscript{5.4}{+2.7}{-1.5}$ & $\tabscript{8.8}{+4.7}{-2.5}$ \\
Observed & 2.6 & 3.4 & 6.5 & 5.4 & 7.9 \\\hline
\multicolumn{6}{c}{${\cal B}\left(H\to {\cal Q}\,\gamma\right)[\tabscript{10}{-3}{}]$} \\ \hline
Expected & $\tabscript{1.2}{+0.6}{-0.3}$ & $\tabscript{1.8}{+0.9}{-0.5}$ & $\tabscript{2.1}{+1.1}{-0.6}$ & $\tabscript{1.8}{+0.9}{-0.5}$ & $\tabscript{2.5}{+1.3}{-0.7}$ \\
Observed & 1.5 &1.3 & 1.9 & 1.3 & 2.0 \\ \hline
\multicolumn{6}{c}{$\sigma\left(pp\to H\right)\times{\cal B}\left(H\to {\cal Q}\,\gamma\right)$ [fb]} \\ \hline
Expected & $\tabscript{26}{+12}{-7}$ & $\tabscript{38}{+19}{-11}$ & $\tabscript{45}{+24}{-13}$ & $\tabscript{38}{+19}{-11}$ & $\tabscript{54}{+27}{-15}$ \\
Observed & 33 & 29 & 41 & 28 & 44 \\ \hline
\hline
\end{tabular}
\end{table}


In conclusion, the first search for the decays of the SM Higgs and $Z$ bosons to
$\Jpsi\,\gamma$ and $\Upsilon(nS)\,\gamma$ ($n=1,2,3$) has been
performed with $\rts=8\TeV$ $pp$ collision data samples corresponding to integrated luminosities of up to $20.3\,\ifb$ collected with the ATLAS detector at the LHC. No
significant excess of events is observed above the background. In the
$\Jpsi\,\gamma$ final state the 95\% CL upper limits on the relevant branching
fractions for the SM Higgs and $Z$ bosons are $1.5\times10^{-3}$ and
$2.6\times10^{-6}$, respectively. The corresponding upper limits in
the $\Upsilon(1S,2S,3S)\,\gamma$ channels are
$(1.3,1.9,1.3)\times10^{-3}$ and $(3.4,6.5,5.4)\times10^{-6}$, for
the SM Higgs and $Z$ bosons respectively. These are the first
experimental bounds on exclusive Higgs and $Z$~boson decays to final states involving
quarkonia.

\vspace{5pt}

We thank CERN for the very successful operation of the LHC, as well as the
support staff from our institutions without whom ATLAS could not be
operated efficiently.

We acknowledge the support of ANPCyT, Argentina; YerPhI, Armenia; ARC,
Australia; BMWFW and FWF, Austria; ANAS, Azerbaijan; SSTC, Belarus; CNPq and FAPESP,
Brazil; NSERC, NRC and CFI, Canada; CERN; CONICYT, Chile; CAS, MOST and NSFC,
China; COLCIENCIAS, Colombia; MSMT CR, MPO CR and VSC CR, Czech Republic;
DNRF, DNSRC and Lundbeck Foundation, Denmark; EPLANET, ERC and NSRF, European Union;
IN2P3-CNRS, CEA-DSM/IRFU, France; GNSF, Georgia; BMBF, DFG, HGF, MPG and AvH
Foundation, Germany; GSRT and NSRF, Greece; ISF, MINERVA, GIF, I-CORE and Benoziyo Center,
Israel; INFN, Italy; MEXT and JSPS, Japan; CNRST, Morocco; FOM and NWO,
Netherlands; BRF and RCN, Norway; MNiSW and NCN, Poland; GRICES and FCT, Portugal; MNE/IFA, Romania; MES of Russia and ROSATOM, Russian Federation; JINR; MSTD,
Serbia; MSSR, Slovakia; ARRS and MIZ\v{S}, Slovenia; DST/NRF, South Africa;
MINECO, Spain; SRC and Wallenberg Foundation, Sweden; SER, SNSF and Cantons of
Bern and Geneva, Switzerland; NSC, Taiwan; TAEK, Turkey; STFC, the Royal
Society and Leverhulme Trust, United Kingdom; DOE and NSF, United States of
America.

The crucial computing support from all WLCG partners is acknowledged
gratefully, in particular from CERN and the ATLAS Tier-1 facilities at
TRIUMF (Canada), NDGF (Denmark, Norway, Sweden), CC-IN2P3 (France),
KIT/GridKA (Germany), INFN-CNAF (Italy), NL-T1 (Netherlands), PIC (Spain),
ASGC (Taiwan), RAL (UK) and BNL (USA) and in the Tier-2 facilities
worldwide.


%

\onecolumngrid
\clearpage 
\begin{flushleft}
{\Large The ATLAS Collaboration}

\bigskip

G.~Aad$^{\rm 85}$,
B.~Abbott$^{\rm 113}$,
J.~Abdallah$^{\rm 152}$,
S.~Abdel~Khalek$^{\rm 117}$,
O.~Abdinov$^{\rm 11}$,
R.~Aben$^{\rm 107}$,
B.~Abi$^{\rm 114}$,
M.~Abolins$^{\rm 90}$,
O.S.~AbouZeid$^{\rm 159}$,
H.~Abramowicz$^{\rm 154}$
% ...
\bigskip
\\
% ...
$^{11}$ Institute of Physics, Azerbaijan Academy of Sciences, Baku, Azerbaijan\\
% ...
$^{85}$ CPPM, Aix-Marseille Universit{\'e} and CNRS/IN2P3, Marseille, France\\
% ...
$^{90}$ Department of Physics and Astronomy, Michigan State University, East Lansing MI, United States of America\\
% ...
$^{107}$ Nikhef National Institute for Subatomic Physics and University of Amsterdam, Amsterdam, Netherlands\\
% ...
$^{113}$ Homer L. Dodge Department of Physics and Astronomy, University of Oklahoma, Norman OK, United States of America\\
$^{114}$ Department of Physics, Oklahoma State University, Stillwater OK, United States of America\\
% ...
$^{117}$ LAL, Universit{\'e} Paris-Sud and CNRS/IN2P3, Orsay, France\\
% ...
$^{152}$ Institute of Physics, Academia Sinica, Taipei, Taiwan\\
% ...
$^{154}$ Raymond and Beverly Sackler School of Physics and Astronomy, Tel Aviv University, Tel Aviv, Israel\\
% ...
$^{159}$ Department of Physics, University of Toronto, Toronto ON, Canada\\
% ...
\end{flushleft}


\end{document}
%